\documentclass[a4paper,12pt]{article}

\usepackage{amssymb,latexsym}

\makeatletter \@addtoreset{equation}{section} \makeatother

\begin{document}

\begin{titlepage}

\thispagestyle{empty}

\begin{flushright}
\hfill{HU-EP-04/18} \\
\hfill{hep-th/0403220}
\end{flushright}

\vspace{35pt}

\begin{center}{ \LARGE{\bf
On supersymmetric solutions  of type \\[4mm]
IIB supergravity with general fluxes}}

\vspace{60pt}

{\bf   Gianguido Dall'Agata}

\vspace{15pt}

{\it  Humboldt Universit\"at zu Berlin,
Institut f\"ur Physik,\\
Newtonstrasse 15, D-12489 Berlin, Germany}\\[1mm] {E-mail:
dallagat@physik.hu-berlin.de}

\vspace{50pt}

{ABSTRACT}

\end{center}

\vspace{20pt}

We propose a general spinor Ansatz to find supersymmetric
configurations preserving 4--dimensional Poincar\'e invariance in the
context of type IIB supergravity in the presence of general fluxes.
We show how this removes the imaginary--selfduality (ISD) constraint on
the 3--form flux and present a simple example with nonvanishing $(0,3)$ flux.
To characterize the geometrical properties of such configurations we
will use the tool of SU(2) structures on the internal space, which 
are naturally linked to the form of the Ansatz we propose.

\end{titlepage}

\newpage

\baselineskip 6 mm

\section{Introduction}

The analysis of supersymmetric backgrounds in the presence of fluxes
has a prominent role in addressing various longstanding problems of
string theory like moduli stabilization.
Vacua preserving 4--dimensional Poincar\'e invariance are especially
interesting because they can be used either in string
compactifications or in the context of the gauge/gravity
correspondence.

In this paper we will focus on ${\cal N} =1$ solutions of type IIB
supergravity and give a geometrical characterization of these type of
backgrounds for generic configurations of fluxes.
In order to do so we are going to use the tool of group structures on
the internal manifold.
This powerful method gives a systematic way to translate supersymmetry
conditions in terms of differential constraints on some structures on
the internal manifold, which then define the metric and flux.
This type of analysis, first introduced in \cite{Gauntlett:2002sc},
has been very fruitful in constructing and classifying new solutions 
in the context of string theory 
\cite{Gurrieri:2002wz}--
\cite{Gauntlett:2004zh} 
(For reviews on string theory solutions with fluxes see also 
\cite{Frey:qqqq,DallAgata:0000}).
It has been especially emphasized in \cite{Martelli:2003ki} that the
choice of an appropriate spinor Ansatz is strictly related to the
possible group structure existing on a manifold and may affect the
form of the resulting solutions.

Looking for ${\cal N} =1$ solutions of type IIB theory, three
different types of supersymmetry parameters were used.
All these assume that there is one globally defined spinor on the
internal manifold $\eta_-$ (with its complex conjugate $\eta_+$) so
that the 10--dimensional supersymmetry parameter $\epsilon$ reads
\begin{eqnarray}
\hbox{Type A: } &\quad& \epsilon = a \, \varepsilon \otimes \eta_- + 
a^* \, \varepsilon^* \otimes \eta_+\,,
\label{eq:Aansatz}\\
\hbox{Type B: } &\quad& \epsilon = a \, \varepsilon \otimes \eta_- \,,
\label{eq:Bansatz}\\
\hbox{Type C: } &\quad& \epsilon =  a \, \varepsilon \otimes \eta_- + b 
\, \varepsilon^* 
\otimes \eta_+\,,
\label{eq:Cansatz}
\end{eqnarray}
with $a$ and $b$ complex functions, $\varepsilon$ is the
4--dimensional supersymmetry parameter and $\eta_{\pm}$ are normalized
to 1.
Type A Ansatz was first introduced by Strominger in the context of
the heterotic theory \cite{Strominger:1986uh}, applied to type IIB
in \cite{Papadopoulos:2000gj} and recently also considered in 
\cite{Kachru:2002sk,Cardoso:2002hd,
Gauntlett:2003cy,Becker:2003yv,Becker:2003sh}.
Type B, named after \cite{Becker:1996gj}, was utilized in the type IIB 
context in \cite{Kehagias:1998gn}--
\cite{Grana:2001xn}.
Finally, a spinor of the form
(\ref{eq:Cansatz}) was proposed in \cite{Frey:2003sd} 
as a way to construct solutions interpolating
between A and B types.
We name it type C, being the third different type of Ansatz used so 
far, though this should not be confused with the S--dual of type A 
which is named in the same way in \cite{Frey:2003sd}.

Based on the existence of a single globally defined spinor,
(\ref{eq:Aansatz})--(\ref{eq:Cansatz}) imply the existence of an
$SU(3)$ structure on the internal manifold.
This structure is characterized by an almost complex structure $J$ and
a 3--form $\Omega$ which are naturally given in terms of $\eta_{\pm}$:
\begin{equation}
J_{mn} = -i \, \eta^{T}_{-} \gamma_{mn}\eta_{+}\,,
\quad \Omega_{mnp} = - \, \eta_+^T \gamma_{mnp} \eta_+\,.
\end{equation}
The supersymmetry conditions following from 
(\ref{eq:Aansatz})--(\ref{eq:Cansatz}) using a metric and a 
5--form preserving Poincar\'e symmetry, impose that the allowed
supersymmetric 3--form fluxes have to be only of type $(2,1)$ or
$(1,2)$ with respect to the natural complex structure $J$.
For type B the requirement is even stronger, because only fluxes which
are of type $(2,1)$ and primitive with respect to $J$ are allowed 
\cite{Giddings:2001yu,Grana:2001xn}.

There are however solutions in the literature which hint to the
possibility of supersymmetric ${\cal N} = 1$ solutions of type IIB
string theory in which the flux may contain also $(3,0)$ and $(0,3)$
contributions \cite{Grana:2000jj,Pilch:2004yg}.
For this reason one should consider a more general Ansatz for the form 
of the supersymmetry parameter which could provide these solutions.
A general Ansatz which gives the desired result is 
\begin{equation}
\hbox{Type D: } \quad \epsilon =  a\, \varepsilon \otimes \eta_- +  \varepsilon^* 
\otimes \left(b\, \eta_+ + c\, \chi_+\right)\,,
\label{eq:Dansatz}
\end{equation}
where a new spinor $\chi$, orthogonal to 
$\eta$, is introduced.
This imposes some strong restriction on the possible choice of internal 
manifold.
One is indeed allowed to employ such an Ansatz only if the internal
manifold has an $SU(2)$ structure and therefore admits 
two globally defined spinors\footnote{It should be remembered that
although the two spinors are orthogonal they are not independent, but,
as we will show later, they are related through the $SU(2)$ structure
tensors.  
This also explains why seemingly different Ans\"atze like in
\cite{Pilch:2004yg} fall instead in our classification.}.
Of course, since an $SU(2)$ structure can be embedded in different 
ways inside an $SU(3)$ structure there is no more the notion of a 
``natural'' complex structure with respect to which one defines the 
Hodge decomposition of forms.
There is actually a $U(1)$--worth of different almost complex structures, 
which therefore can be parameterized by a phase.
Anyway, once a choice is made, one can see that it is possible to 
combine $(3,0)$ and/or $(0,3)$ fluxes with supersymmetry.
Also the statements about the integrability of such a structure will 
depend on the phase choice.

Let us stress here that although we look for ${\cal N} = 1$ solutions, 
the configurations we obtain may preserve more supersymmetry.
Already in the case of a strict $SU(3)$ structure, namely an internal manifold 
allowing for just one globally defined spinor, 
supersymmetric configurations may preserve ${\cal N} = 2$.
This happens for instance in the degenerate case of zero fluxes, where 
the group structure is identified with the holonomy of the internal 
manifold, which is then Calabi--Yau.
This, of course, is even more true in the case of $SU(2)$ structures.

It is interesting to notice that, due to the $SU(2)$ structure,
the internal manifold will always admit an almost product structure
which allows us to discuss them as fibrations of 2--manifolds on
4--manifolds.
In simple cases this structure is even integrable, therefore
simplifying the analysis and allowing to exhibit solutions in closed
form.
In this paper we will discuss the form of the SU(2) structures allowed 
by supersymmetry when the type D Ansatz (\ref{eq:Dansatz}) is used for 
the spinor parameter.
We will show that the generic internal manifold may be non--complex and 
indeed allows for (3,0) or (0,3) fluxes.
As an application we will provide a simple solution with $(0,3)$ flux
and holomorphic dilaton.
This is a warped product of Minkowski space--time with an internal 
manifold which is an $\mathbb{R}^2$ fibration over a $K3$.

After this introduction, in section \ref{susystruc} we will discuss in 
detail the spinor Ansatz and its relation to both $SU(3)$ and $SU(2)$ 
structures.
Then, in section \ref{susyanalis}, we will perform a detailed analysis
of the supersymmetry conditions reinterpreting them in terms of the
$SU(2)$ structures.
Finally, we will show how to produce simple solutions with $(3,0)$ and
$(0,3)$ fluxes in section \ref{solutions} and conclude with some 
comments in section \ref{comments}.
We also give an appendix with more details on conventions, 
notations and $SU(2)$ structures in 6 dimensions.

\section{Supersymmetry Ansatz and group structures}
\label{susystruc}

\subsection{Preliminaries}

Type IIB supersymmetry transformations \cite{Schwarz:1983wa,Schwarz:1983qr} read
\begin{eqnarray}
\delta \psi_M & = & \frac1\kappa \left( D_M - 
\frac{i}{2}Q_M\right)\epsilon+ \frac{i}{480} \, \Gamma^{M_1 \ldots M_5} F_{M_1\ldots 
M_5} \gamma_M \epsilon - \frac{1}{16} \, \Gamma_M G \epsilon^* - \frac18 
\, G \, \Gamma_M \epsilon^* \,, \label{eq:gravitino}\\
\delta \lambda & = & \frac{i}{\kappa} \Gamma^M P_M \epsilon^* - 
\frac{i}{4} G \epsilon\,, \label{eq:dilatino}
\end{eqnarray}
where we used the contraction $G \equiv 1/6 \, G_{MNP} \Gamma^{MNP}$.
Here and in the following we will use the conventions of
\cite{Schwarz:1983qr}, where the supersymmetry conditions and
covariant equations of motion for type IIB supergravity were first
derived\footnote{For a discussion and derivation of a covariant action
see \cite{DallAgata:1997ju,DallAgata:1998va}.}.
The definition of the dilaton--axion curvature $P_{M} = f^2 \partial_M
B$ and $U(1)$ connection $Q_M = f^2 \hbox{Im }\left(B \partial_M
B^*\right)$ can be related to the standard stringy
quantities $\tau = C + i \, e^{-\phi}$ by using $B = \frac{1+ i
\tau}{1-i \tau}$ and $f^2 = 1/(1- |B|^2)$.
The 3--form field strength $G = f\left(F_3 - B F_3^*\right)$, where
$F_3 = d A_2$, follows from a complex 2--form $A_2$ which is related
to the usual string NS and RR forms as $\kappa A_2 = g \left( B_{NS} +
i B_{RR}\right)$.
The Bianchi identities for these forms are
\begin{eqnarray}
dP & = & 2 i \, Q \wedge P\,, \label{eq:dP}\\
dG & = &  i\, Q \wedge G + \overline G \wedge P\,, \label{eq:dG}\\
dF_5 & = &  i\, \frac{\kappa}{8} \,  G \wedge\overline G\,, \label{eq:dF5}
\end{eqnarray}
together with the connection which satisfies $dQ = i \,\overline P \wedge P$.
All the spinors are complex Weyl spinors satisfying 
$\Gamma^{11} \psi_M = - \psi_M$, $\Gamma^{11} \lambda = \lambda$ and 
$\Gamma^{11}\epsilon = -\epsilon$.

As said in the introduction, we are looking for solutions which 
preserve 4--dimensional Poincar\'e invariance.
This type of solutions can be used for compactifications of 
type IIB supergravity as well as for the gauge/gravity 
correspondence.
It should indeed be noticed that $AdS_5$ can be written as an 
$\mathbb{R}$ foliation of 4--dimensional Minkowski spacetime and that the 
domain--wall solutions describing holographic renormalization group 
flows are also warped products of 4--dimensional Minkowski space with 
$\mathbb{R}$ and some 5--dimensional compact space which, together, define a non 
compact 6--dimensional internal manifold.
For the 10--dimensional metric we will then take a  
warped product of 4--dimensional Minkowski space and some internal 
6--dimensional euclidean manifold \cite{Grana:2001xn}
\begin{equation}
ds^2 = \frac{1}{\sqrt{Z(y)}} \,\eta_{\mu\nu}\, dx^\mu \otimes dx^\nu + 
\sqrt{Z(y)} \, g_{mn}(y) \, dy^m \otimes dy^n\,.
\label{eq:metric}
\end{equation}
In order to preserve 4--dimensional Poincar\'e invariance on the full 
solution we also ask that the RR 5--form satisfies
\begin{equation}
F_{0123m} = \partial_m h(y)\,,
\label{eq:F5ansatz}
\end{equation}
as well as the self--duality property $*F_5 = F_5$.
We do not impose any restriction on the dilaton or on the 3--form flux.

Before solving the supersymmetry conditions one needs to further 
specify an Ansatz for the spinor parameter and therefore we will now 
discuss more in detail how we arrived to the (\ref{eq:Dansatz}) Ansatz.

\subsection{Spinor Ansatz}

Since we are looking for solutions preserving 4--dimensional 
Poincar\'e invariance, we will use this property to perform a ``4+6'' 
splitting of the supersymmetry spinor as well as of the 10--dimensional 
$\Gamma$ matrices.
For generic ${\cal N} = 1$ solutions this implies that the 10--dimensional 
supersymmetry parameter gets factorized
\begin{equation}
\epsilon(x,y) = \varepsilon(x) \otimes \zeta^1(y) + 
\varepsilon^*(x)\otimes (\zeta^2(y))^*\,,
\label{eq:N1spinor}
\end{equation}
where $\varepsilon$ is a Weyl spinor on Minkowski space--time and
$\zeta^1$, $\zeta^2$ are generic 6--dimensional chiral spinors on the
internal manifold.

In order to explicitly solve the supersymmetry equations it is 
important to establish the properties and relations between these 
spinors.
First one should note that due to the chirality properties of the 10-- 
and 4--dimensional spinors $\Gamma^{11}\epsilon(x,y) = -\epsilon(x,y)$, 
$\gamma^5\varepsilon(x) = \varepsilon(x)$ and the definition of the 
10--dimensional $\Gamma$ matrices in terms of the lower dimensional 
ones (see the appendix for more details), $\zeta^i$ should have a 
definite chirality $\gamma^{7}\zeta^i=-\zeta^i$.
More restrictions can then follow by assuming that these 2 spinors are 
not unrelated, as in \cite{Papadopoulos:2000gj,Grana:2001xn,Frey:2003sd}.
Generically, a globally well defined supersymmetric solution implies 
that the aforementioned spinors are also globally defined and, 
accordingly, that the structure group of the tangent bundle over the 
internal manifold is reduced.
The group of transformations $G$ required to patch the tangent bundle over 
the manifold is not the generic $SO(6)$ group of a Riemannian six--manifold, 
but a subgroup of it, $G \subset SO(6)$.
The spinors which are globally defined, those which define our supersymmetry 
parameters, must not transform under $G$ and therefore are singlets under 
the $SO(6) \to G$ decomposition.
When $\zeta^1$ and $\zeta^2$ are chosen parallel, or one of the two is 
vanishing, then the internal manifold displays an $SU(3)$ structure.
Decomposing $SO(6)\simeq SU(4) \to SU(3)$, we see that the spinorial
representation admits one singlet $4 \to 3 + 1$.
This implies that an $SU(3)$ structure admits just one globally defined
complex chiral spinor which is to be identified with the surviving
supersymmetry parameter $\zeta$.
Since this latter is globally defined, we can always normalize it to
1, or extract from it the normalized spinor, say $\eta_-$.
The subscript refers to the 6--dimensional chirality $\gamma^7
\eta_{\pm} = \pm \eta_{\pm}$.
The general form of the supersymmetry parameter for solutions with an 
internal manifold preserving an $SU(3)$ structure is then 
\begin{equation}
\epsilon(x,y) = \varepsilon(x) \otimes a(y)\, \eta_-(y) 
 + \varepsilon^*(x) \otimes b(y) \,\eta_+(y)\,, 
\label{eq:eps0}
\end{equation}
where $a,b$ are complex functions and $\eta_+ = \eta_-^*$.
This includes all the solutions presented in the paper \cite{Frey:2003sd} 
and more because we allow for the norms of $a$ and $b$ to be unrelated.
For $a = b^*$ we recover the type A Ansatz (\ref{eq:Aansatz}) and 
for $b=0$ we recover the B Ansatz (\ref{eq:Bansatz}).
The functions $a$ and $b$ need not be related, though.
The 10--dimensional supersymmetry parameter is a Weyl spinor, which can 
be written as the sum of 2 real Majorana--Weyl spinors: $\epsilon = 
\epsilon_1 + i \,\epsilon_2$.
Having an $SU(3)$ structure means that we can write them as
\begin{equation}
\epsilon_i = \varepsilon \otimes f_i\, \eta_- + \varepsilon^* \otimes 
f_i^*\, \eta_+\,,
\label{eq:10MW}
\end{equation}
where $f_i(y)$ are arbitrary complex functions.
This finally implies that $a = f_1 + i f_2$ and $b = f_1^* + i f_2^*$ 
which therefore are generically unrelated complex functions too.

The $SU(3)$ structure can also be described \cite{chiossi} by an almost 
complex structure $J$ and a globally defined 3--form $\Omega$.
These are also singlets under the $SO(6)\to SU(3)$ decomposition and 
arise in the product of the fundamental invariant spinors:
\begin{eqnarray}
J_{mn} &=& -i \, \eta^{T}_{-} \gamma_{mn}\eta_{+}\,,
\label{eq:Jdef}\\
\Omega_{mnp} &=& - \, \eta_+^T \gamma_{mnp} \eta_+\,.
\label{eq:psidef}
\end{eqnarray}
By using these tensors in the analysis of the supersymmetry
conditions, once contracted with $\eta_{\pm}$, one can extract quite
easily the conditions
\begin{equation}
\Omega \wedge G = 0 = \overline{\Omega} \wedge G\,.
\label{eq:niente30}
\end{equation}
This means that having an $SU(3)$ structure on the internal space,
which forces the (\ref{eq:eps0}) spinor Ansatz, one can only use
3--form fluxes which are of $(2,1) + (1,2)$ type with respect to the
natural complex structure defined by the covariantly constant spinor
$\eta_{\pm}$.
Any $(3,0)$ or $(0,3)$ flux with respect to $J$ defined by (\ref{eq:Jdef}) 
breaks supersymmetry.

A possible wayout is to make a stronger requirement on the structure 
of the internal manifold and ask for an $SU(2)$ structure.
The existence of an $SU(2)$ structure on the internal manifold implies
the possibility of having an additional globally defined spinor
$\chi$, which may be used to modify (\ref{eq:eps0}).
There are indeed 2 singlets in the decomposition of the fundamental 
representation of $SU(4)$ under $SU(2)$.
Taking again $\chi$ to be canonically normalized and using the same 
conventions as for $\eta$, 
we can generically write the 10--dimensional spinor Ansatz as
\begin{equation}
\epsilon(x,y) = \varepsilon(x) \otimes \left[ f_1(y) \, \eta_-(y) + 
f_2\, (y) \chi_-(y)\right] + \varepsilon^*(x) \otimes \left[ f_3(y)\, \eta_+(y) + 
f_4(y)\, \chi_+(y)\right]\,,
\label{eq:eps1}
\end{equation}
with $f_i\in {\mathbb C}$ generic functions to be determined.
It is useful to notice however that one can always simplify the above 
Ansatz, by removing one of the functions by a field redefinition.
This is easily proved if one notices that $\phi_- \equiv 
\frac{f_1 \eta_- + f_2 \chi_-}{|f_1|^2 + |f_2|^2}$ has norm one and defines an 
SU(2) structure together with the orthogonal combination 
$\psi_- \equiv 
\frac{f_2^* \eta_- - f_1^* \chi_-}{|f_1|^2 + |f_2|^2}$.
We remind that for the spinor to be well--behaved everywhere on the solution, we also 
have to require that the $f_i$ functions are globally defined.

We have therefore argued that the most general spinor Ansatz for an 
$SU(2)$ structure is the Ansatz D (\ref{eq:Dansatz}) given in the introduction:
\begin{equation}
\epsilon(x,y) =  a(y)\, \varepsilon(x) \otimes \eta_-(y) +  
\varepsilon^*(x) \otimes \left(b(y)\, \eta_+(y) + c(y)\, \chi_+(y)\right)\,.
\end{equation}

One could think about having even more restricting structures on the 
tangent bundle, of course.
The result however would be that the internal space factorizes, often 
leading to solutions preserving more supersymmetries, and we 
will therefore not consider it further here.

\subsection{SU(3) and SU(2) structures}

We have just seen how the existence of different group structures 
allows for different spinor Ans\"atze which may lead to new 
supersymmetric solutions.
Since in what follows we are going to make an extensive use of them, 
we will give some more details on the introduction of a 
$G$--structure, for $G = SU(3)$ or $G = SU(2)$.
Some parts of what follows are also covered in 
\cite{Gauntlett:2003cy,DallAgata:2003ir,Gauntlett:2004zh}.

In the previous section we have learned how globally defined spinors imply
a group structure on the tangent bundle of a 6--manifold.
Alternatively, one can characterize a $G$--structure by invariant forms, 
i.e. tensors which are singlets in the decomposition under $G$.
For an $SU(3)$ structure one needs a 2--form $J$ and a complex
3--form $\Omega$, as defined in (\ref{eq:Jdef}), (\ref{eq:psidef}), which 
satisfy the compatibility constraints
\begin{equation}
J \wedge \Omega = 0\,, 
\qquad J \wedge J \wedge J = \frac34 i\, \Omega \wedge \overline 
\Omega\,.
\label{eq:compa}
\end{equation}
An $SU(2)$ structure requires the existence of a triplet of 2--forms
$J^i$ and a complex 1--form $w$.
The 1--form must lie on an orthogonal space with respect to the one 
spanned by the 2--forms
\begin{equation}
w\, \lrcorner \,J^i = 0\,
\label{eq:ocdn}
\end{equation}
and these latter give the volume of a 4--dimensional subspace 
\begin{equation}
J^i \wedge J^j = 0\,, \hbox{ for } i \neq j\,, \qquad J^i \wedge J^i = 
Vol_4\,.
\label{eq:vol4}
\end{equation}
One can also use the metric to raise one of the indices of the 2--forms 
and obtain a triplet of almost complex structures on the base 
satisfying 
\begin{equation}
J^i J^j = -\delta^{ij}-\epsilon^{ijk}J^k\,.
\label{eq:tripl}
\end{equation}
These can also be combined again into a 2--form, a complex 3--form and
a complex 1--form.

As for the $SU(3)$ structures, one can build the $SU(2)$ structure 
tensors from combinations of the invariant spinors.
We could provide directly the expression analogous to (\ref{eq:Jdef})
and (\ref{eq:psidef}), but it is instructive to derive the $SU(2)$ 
tensors starting from an existing $SU(3)$ structure and imposing the
existence of some additional independent spinor.
In this way it will be clear the difference between the previous 
analyses of the supersymmetry conditions and the one presented here. 

With respect to $SU(3)$ structures, the $SU(2)$ ones are characterized
by the existence of an additional globally defined complex vector $w$.
One can therefore construct them from existing $SU(3)$ structures by
introducing new globally defined spinors as
\begin{equation}
\chi_{+} = \frac12 \,w_m \gamma^m \eta_-\,, \quad \chi_- = \frac12\, 
\overline{w}_m \gamma^m \eta_+\,,
\label{eq:chisp}
\end{equation}
where it is clear that $\chi^{*}_{\pm} = \chi_{\mp}$.
Assuming that the $\eta_\pm$ spinors and the globally defined vector 
$w$ are canonically normalized, i.e. $\eta_{\pm}^{\dagger} 
\eta_{\pm} =1$ and $w_m \overline w^m = -2$, 
one obtains more orthogonality relations
\begin{equation}
\chi_{\pm}^\dagger \chi_{\pm} = 1\,, \quad
\chi_{\pm}^\dagger \eta_{\pm} = 0\,,
\label{eq:orthSU2}
\end{equation}
where we also used that $w$ is holomorphic with respect to the almost 
complex structure defined by (\ref{eq:Jdef}) ${J_m}^n w_n = i \, w_m$.

Combining the information coming from the definition of the $SU(3)$ 
structure (\ref{eq:Jdef}), (\ref{eq:psidef}), the definition of the $SU(2)$ 
spinors (\ref{eq:chisp}) and the orthogonality properties (\ref{eq:orthSU2}) 
we can then obtain the definitions of all the invariant tensors in 
terms of the fundamental spinors:
\begin{equation}
\begin{array}{rclcrcl}
\eta_+^T \gamma^m \chi_+ &=& w^m\,, &\quad&
\eta_-^T\gamma_{mn}\eta_+ &=& i \,J_{mn}\,, \\[2mm]
\eta_+^T\gamma_{mn}\chi_- &=& {K}_{mn}\,, &\quad&
\eta_-^T\gamma_{mn}\chi_+ &=& \overline{K}_{mn}\,,\\[2mm]
\chi_-^T \gamma_{mn}\chi_+ &=& 2\, w_{[m}\overline{w}_{n]} - i \,
J_{mn}\,, &\quad&
\eta_+^T\gamma_{mnp}\eta_+ &=& -\Omega_{mnp}\,,
\end{array}
\label{eq:su2tens}
\end{equation}
where we have introduced the combination $K \equiv J^2 + i J^3$.
Moreover, using the identities on the 6--dimensional $\gamma$ matrices and the
spinors $\eta$ and $\chi$ reported in the appendix, it can also be
checked that $J = J^1 - \frac{i}{2}\, w \wedge \overline w$ and
$\Omega = K \wedge w$.

In doing so, we have made a choice for the embedding of the $SU(2)$ 
structure inside the $SU(3)$ one.
This was done in a somewhat natural way because we have constructed 
the $SU(2)$ structure starting from an existing $SU(3)$ one.
However, it should be clear that given 2 globally defined spinors 
there is no preferred choice for the one describing the $SU(3)$ 
structure.
Let us then establish how the $SU(2)$ structures are embedded into 
the $SU(3)$ ones generically.
Considering always normalized spinors, there is a $U(1)$ degeneracy of possible 
spinors defining the $SU(3)$ structure starting from the 2 globally 
defined spinors describing the $SU(2)$ one:
\begin{equation}
\beta_- = \cos \phi \, \eta_- + \sin \phi \, \chi_-\,.
\label{eq:beta}
\end{equation}
Using this normalized spinor one can then define the $SU(3)$ structure 
as given in (\ref{eq:Jdef}) and (\ref{eq:psidef}), now with $\beta_-$ 
replacing $\eta_-$:
\begin{eqnarray}
J &=& \left(1- 2 \sin^2 \phi\right) J^1  - \frac{i}{2}w \wedge \overline 
w + \frac{i}{2}\, \sin 2\phi\, (K - \overline K)\,, \label{eq:Jsu23}\\
\Omega &=& \cos^2 \phi\, K \wedge w
+ \sin^2\phi \, \overline{K} \wedge w + i \, \sin 2\phi\, J^1 \wedge w\,.
\label{eq:su3su2}
\end{eqnarray}
It is now clear that the properties of the $SU(3)$ structure depend on 
the choice of the embedding angle.
Among these, the integrability of the complex structure.
Although this will not change the fact that the solution be 
supersymmetric or not, it may change the physical interpretation as
there are instances in which there is a natural choice of complex 
structure which specifies $\phi$.
In the following we will consider for definiteness $\phi=0$, i.e. 
$\beta_- = \eta_-$ as done previously.

\section{Analysis of supersymmetry conditions}
\label{susyanalis}

In this section we are going to interpret the supersymmetry equations 
as conditions on the intrinsic torsion of the internal manifold.
This means that we are going to specify some differential constraints 
on the $SU(2)$ structure tensors defined in (\ref{eq:su2tens}).
At the same time we will show that there are some constraints on the 
3--form fluxes and dilaton/axion, though they 
will not be as restrictive as those obtained previously using the 
Ans\"atze (\ref{eq:Aansatz})--(\ref{eq:Cansatz}).
For this reason, we list here the general expansion of the
fluxes in terms of the $SU(2)$ structures, so that statements on the 
single modules can be made more precise.
The dilaton/axion can be decomposed in 3 pieces
\begin{equation}
P  =  p_1 \, w + p_2 \,\overline w + \Pi\,, \label{eq:Pfl}
\end{equation}
where $w \, \lrcorner \, \Pi = 0$, for a real $\Pi$.
The 3--form flux is then
\begin{eqnarray}
G  &=&  g_{30}\, K \wedge w +g_{21}\, K \wedge \overline{w} + 
\tilde{g}_{21} \,J^1 \wedge w + J^1 \wedge V_1 + w \wedge \overline{w} 
\wedge V_2+ \nonumber\\
&+& w \wedge T_1 + \overline{w} \wedge T_2 + 
g_{12}\,\overline{K}\wedge w + \tilde{g}_{12}\,J^1 \wedge \overline{w}
+  g_{03}\, \overline{K}\wedge \overline{w} \,,
\label{eq:Gflux}
\end{eqnarray}
where the flux components further satisfy
\begin{equation}
J^i \wedge T_{1,2} = 0\,, \quad w \, \lrcorner \;T_{1,2} = 0\, \quad 
\hbox{and}\quad   w\, \lrcorner\; V_{1,2} = 0\,.
\label{eq:constrT}
\end{equation}
We can also combine $V_1$ and $V_2$ so that the primitive and 
non--primitive parts with respect to $J$ are explicit:
$ J^1 \wedge V_1 + w \wedge \overline{w} 
\wedge V_2 = \frac12 \, J \wedge \left(V_1 + 2i \, V_2\right) + 
\frac12\, \left(J^1 + \frac{i}{2}\,w \wedge \overline w\right)\wedge 
\left(V_1 - 2i \, V_2\right)$.
We also notice that $g_{12}$, $g_{21}$ refer to primitive fluxes with 
respect to $J$ and $\tilde g_{21}$, $\tilde g_{12}$ to non--primitive 
ones.
Restriction to ISD fluxes can be made by setting
\begin{equation}
g_{30}=g_{12}=\tilde g_{21} = 0\,,\quad T_2 = 0\,, 
\quad (1+iJ)(V_1 -2 i\, V_2) = 0 = (1-iJ)(V_1+2i \,V_2)\,.
\label{eq:ISD}
\end{equation}

The supersymmetry equations (\ref{eq:gravitino}), (\ref{eq:dilatino}) are 
not written in this language.
In order to extract the information we want in terms of $SU(2)$
structure conditions we should perform a projection on
the full basis of independent spinors and use the relations between 
the spinor bilinears and the $SU(2)$ structure tensors (\ref{eq:su2tens}).
For the case at hand, one should project on the full basis given by 
$\eta^\dagger_{\pm}$, $\eta^\dagger_{\pm} \gamma^{m\ldots}$, up to 3 $\gamma$ 
matrices, and the same for the spinors $\chi_{\pm}$.
Luckily, it is not necessary to consider all these projections, because only 
a subset of them gives independent conditions.
As a first fact one can notice that only the projections along 
$\eta_{\pm}$ are necessary due to the relation (\ref{eq:chisp}) between 
$\eta_{\pm}$ and $\chi_\pm$.
Then, using group theory, from the $SU(4) \to SU(3) \to SU(2)$
decomposition one learns that the only independent objects one can
build are $\eta_{\pm}$ and $\gamma^{m}\eta_{\pm}$.
Therefore we will just consider these projections.

We will start analyzing first the dilatino equation 
(\ref{eq:dilatino}) and then the gravitino equation 
(\ref{eq:gravitino}) for the free index in the external space and then 
in the internal space.
The first two sets of equations will give us conditions on the fluxes as 
well as determine the warp--factor and 5--form flux in terms of the 
3--form flux.
Then, from the internal gravitino we will get the proper differential 
constraints on the $SU(2)$ structure.

Let us start with the analysis of the dilatino equation.
Making explicit use of the D spinor Ansatz, (\ref{eq:dilatino}) reads
\begin{eqnarray}
\delta \lambda = 0&\Rightarrow &\gamma^m P_m \left[ - \varepsilon^* \otimes 
\left(a^* \, \eta_+ \right) 
+ \varepsilon \otimes \left( b^*\, \eta_- + c^* \,\chi_-\right) 
\right] = \nonumber \\
&=&\frac{\kappa}{24 \sqrt{Z}} \gamma^{mnp}\left[\varepsilon 
\otimes \left( a \,\eta_- \right) - \varepsilon^* \otimes 
\left(b \,\eta_+ + c \,\chi_+\right)\right] G_{mnp}\,.
\label{eq:dilat1}
\end{eqnarray}
Simple contractions of this equation with $\eta_{\pm}^T$ and 
$\eta_\pm^T \gamma_m$ give us 
two constraints on the fluxes
\begin{eqnarray}
2 b\, g_{03} +i c \, \tilde{g}_{12} &=& 0\,,
\label{eq:constdil1}\\
2 b^*\, g_{30} -i c^* \, \tilde{g}_{21}&=& 0\,,
\label{eq:constdil2}
\end{eqnarray}
and equations which determine the dilaton/axion in terms of the 
components of the 3--form flux as specified in (\ref{eq:Gflux})
\begin{eqnarray}
\sqrt{Z} c^* \, p_1 & = & - \kappa a \,g_{30}\,,
\label{eq:v1}\\
2 \sqrt{Z} a^* \, p_2 & = & \kappa \, \left(2 c \, g_{21}+ i b \,
\tilde{g}_{12}\right)\,, \label{eq:v2}\\
a^*\, \left(1 + i J\right) \Pi & = & 
i \,\frac{\kappa}{4\sqrt{Z}}\left[b\, \left(1+i J\right)
\left(V_1+2i \,V_2\right)+\right. \label{eq:v4}\\
&-& \left. c\, \overline{K} \left(V_1 - 2i 
\,V_2\right)\right]\,,\nonumber \\
b^*\, \left(1-iJ\right) \Pi+ c^*\, K\,
\Pi& = & -\frac{\kappa}{4\sqrt{Z}} \, i a \,
\left(1-iJ\right) \left(V_1+2i\, V_2\right)\,. 
\label{eq:v5}
\end{eqnarray}
From now on, when $J$, $J^1$ or $K$ are not separated by wedge 
products from other forms we understand the associated structures, 
i.e. $J \, V\equiv e^a\, {J_a}^b\, V_b$.
It should be noted from (\ref{eq:v1}), (\ref{eq:v2}),using 
(\ref{eq:constdil1}), that the $(3,0)$ 
and $(0,3)$ components of the 3--form flux drive a holomorphic and 
anti--holomorphic part of the dilaton, respectively.
Moreover, as noted also in \cite{Grana:2000jj}, the appearance of a 
$(0,3)$ flux must be accompanied by a non--primitive $(1,2)$ component.


As a second step one has to analyze the gravitino equation 
(\ref{eq:gravitino}), when the free index is on the external space 
$M = \mu$.
Using that the 4--dimensional spinor on Minkowski space can be chosen 
to be constant $\partial_\mu \varepsilon(x) = 0$, 
the gravitino equation reads
\begin{eqnarray}
\delta \psi_{\mu}= &-&\frac{1}{8\kappa\sqrt{Z}}\, \gamma^m\partial_{m}\log Z\left[ 
\gamma_\mu \varepsilon \otimes \left(a\, \eta_- \right) - 
\gamma_\mu \varepsilon^* \otimes \left(b\, \eta_+ + c\,
\chi_+\right)\right] +\nonumber\\
&+& \frac{1}{2} \sqrt{Z}\, \gamma^m\partial_{m}h \left[ 
\gamma_\mu \varepsilon \otimes \left(a \,\eta_- \right) +
\gamma_\mu \varepsilon^* \otimes \left(b \,\eta_+ + c\,
\chi_+\right)\right] +
\label{eq:grav1}\\
&+& \frac{1}{16\cdot 6 Z} \gamma^{mnp}G_{mnp} \left[ 
-\gamma_\mu \varepsilon^* \otimes \left(a^* \,\eta_+ \right) +
\gamma_\mu \varepsilon \otimes \left(b^* \,\eta_- + c^* \,
\chi_-\right)\right] =0\,, \nonumber
\end{eqnarray}
where we also made explicit the connection of the warped metric, 
expressing everything in terms of the quantities on the unwarped 
spaces.

Again, projecting along $\eta_\pm^T$ and $\eta_\pm^T \gamma_m$, we 
obtain further constraints on the fluxes and determine  
the warp factor and 5--form flux in terms of the 3--form flux.
We also decompose the derivative of the warp factor and the 5--form flux 
to make explicit the irreducible $SU(2)$ modules 
\begin{eqnarray}
\partial_m \log Z &=&  \sigma \, w_m + \overline \sigma \, \overline w_m + 
\Sigma_m\,,
\label{eq:logZ}\\
\partial_m h &=& \theta \, w_m + \overline \theta\, \overline w_m + H_m\,,
\label{eq:dmh}
\end{eqnarray}
where $w \, \lrcorner\; \Sigma = w\, \lrcorner\; H = 0$ and $\Sigma$ 
and $H$ are real.
Plugging these expressions into (\ref{eq:grav1}) and performing the
appropriate projections, not only we get the same flux constraint as
(\ref{eq:constdil2}), but also some equations determining the warp
factor and the 5--form flux
\begin{eqnarray}
\sigma - 4\kappa Z \, \theta & = & \frac{\kappa}{\sqrt{Z}}\,
\left(2 \frac{c^*}{a}\, g_{12} - i \frac{b^*}{a}\, \tilde g_{21}\right)\,, 
\label{eq:gg1}\\
\overline \sigma + 4 \kappa Z \,\overline \theta & = & -\frac{2\kappa 
a^*}{\sqrt{Z} c} \, g_{03}\,,
\label{eq:gg2}
\end{eqnarray}
\begin{eqnarray}
\left(1-iJ\right)\left[\Sigma - 4\kappa Z \, 
H\right]    &=& - i\frac{b^* \kappa}{2 
a\sqrt{Z}}\, \left(1-iJ\right)\left(V_1 +2i \, 
V_2\right) + \nonumber \\
&+&i \frac{c^* \kappa}{2 a \sqrt{Z}}\, {K}\left(V_1 -2i \,
V_2\right)\,,
\label{eq:gg3}\\
\left(1+iJ\right)\left[\Sigma + 4\kappa Z \, 
H\right]   &+& \frac{c}{b}\, \overline{K}\left(\Sigma + 4\kappa Z \, 
H\right)  =\nonumber\\
&=&  i\frac{a^* \kappa}{2 b\sqrt{Z}}\, \left(1+iJ\right)\left(V_1 +2i \, 
V_2\right)\,.  \label{eq:gg4}
\end{eqnarray}
It should be noted that restriction to ISD fluxes imposes a precise
relation between the warp--factor and 5--form flux.
Indeed, using (\ref{eq:ISD}), the right hand side of both
(\ref{eq:gg1}) and (\ref{eq:gg3}) vanishes and therefore we get that 
$d\log Z = 4 \kappa Z\, dh$.
This relation, using different conventions, was pointed out in
\cite{Gubser:2000vg,Giddings:2001yu}.


The last, and more demanding, step is given by computing the 
differential constraints on the $SU(2)$ structure from the analysis of 
the variation of the internal gravitino $\delta \psi_m$.
Since the various structures (\ref{eq:su2tens}) are defined in terms
of $\eta_\pm$ and $\chi_\pm$, in order to obtain differential
constraints on them we have to extract $\nabla_m \eta_\pm$ and
$\nabla_m \chi_{\pm}$ from $\delta\psi_m = 0$.
This can be done by separating the contributions proportional to 
$\varepsilon(x)$ from those proportional to $\varepsilon^*(x)$ and, in 
the case of $\nabla \chi_-$, by taking appropriate linear combinations 
of them.
The resulting expressions are 
\begin{eqnarray}
\nabla_m \eta_- &=& 
\left(-\partial_m \log a  + \frac{i}{2}\, Q_m \right) \eta_- - \frac{1}{8}{\gamma_m}^n \partial_n \log Z \,\eta_- -
\frac{\kappa}{2}\, \gamma^n\gamma_m \,
Z\, \partial_n h \,\eta_-  +\nonumber \\
&+& \frac{\kappa}{\sqrt{Z}} \, \frac{b^*}{a}\, G_{pqr} 
\left(\frac{1}{16\cdot 6}{\gamma^{pqr}}_m + 
\frac{3}{32}\,\gamma^{pq}\delta^{r}_m\right)\eta_- + 
\label{eq:nablaeta}\\
&+& \frac{\kappa}{\sqrt{Z}} \frac{c^*}{a}G_{pqr}\left(\frac{1}{16 
\cdot  6}{\gamma^{pqr}}_m + 
\frac{3}{32}\,\gamma^{pq}\delta^{r}_m\right)\chi_-\,,
\nonumber
\end{eqnarray}
and 
\begin{eqnarray}
\nabla_m \chi_- &=& -\left( \partial_m \log c^* + \frac{i}{2}\, 
Q_m \right)\chi_- + \left(
\frac{b^*}{c^*}\partial_m \log a - \frac{1}{c^*}\, \partial_m b^* - 
i\frac{b^*}{c^*} Q_m \right)\eta_- + \nonumber\\
&-& \frac{1}{8}{\gamma_m}^n \, \partial_n \log Z \,\chi_- + 
\frac{\kappa}{2}\gamma^n\gamma_m\, 
Z\, \partial_n h \,\chi_- + \kappa \,\frac{b^*}{c^*}\, \gamma^n\gamma_m \,
Z\, \partial_n h \,\eta_- \nonumber \\
&-& \frac{\kappa}{\sqrt{Z}} \frac{b^*}{a} \,G_{pqr}\, 
\left(\frac{1}{16 \cdot  6}{\gamma^{pqr}}_m + 
\frac{3}{32}\,\gamma^{pq}\delta^{r}_m\right)\chi_- + 
\label{eq:nablachi}\\
&-& \frac{\kappa}{\sqrt{Z}} \frac{(b^*)^2 G_{pqr} - a^2 \overline{G}_{pqr}}{c^*a}
\left(\frac{1}{16 \cdot  6}{\gamma^{pqr}}_m + 
\frac{3}{32}\,\gamma^{pq}\delta^{r}_m\right)\eta_-\,.
\nonumber
\end{eqnarray}

We can now compute $dw$ and $dJ^i$, but from the orthogonality 
properties of the $\eta$ and $\chi$ spinors we can also obtain some 
differential equations specifying the behaviour of the norms $a$, $b$ 
and $c$ appearing in (\ref{eq:Dansatz}) in terms of the fluxes.
First of all, by recalling that $d(\eta_+ \eta_-) = 0 =d(\chi_+ \chi_-) $ 
we get differential conditions on the absolute value of the $a$ and 
$c$ functions:
\begin{eqnarray}
2 d \log |a| &=& -\kappa Z \, d h -\left\{\frac{\kappa}{4\sqrt{Z}}w
\left(i \frac{b^*}{a}\tilde g_{21} -2i 
\frac{b}{a^*} \tilde g_{12}^* - 2 \frac{c^*}{a}g_{12}- 4 
\frac{c}{a^*}g_{03}^*\right) 
+\right. \nonumber \\
&-& \frac{\kappa}{16 \sqrt{Z}}\left[ 
\frac{b^*}{a}\left(J - 3 i\right)  \left(V_{1} +
2i\, V_{2}\right)-
\frac{c^*}{a}\, K  \left(2{V_2} -3 i \,
{V_1}\right)\right] + \nonumber \\
&+& \left.c.c.\right\} \,,
\label{eq:da}
\end{eqnarray}
\begin{eqnarray}
2 d \log |c| &=& \kappa Z \, d h + \kappa Z 
 \left(\frac{b^*}{c^*} \, \overline{K}+ \frac{b}{c} \, K \right)   dh 
+ \nonumber \\
&-&\left\{ \frac{\kappa}{16 \sqrt{Z}}\left[ 
\frac{b^*}{a}\left(J + 3 i \right)  \left(V_1 -
2i \,V_2\right)+  
\frac{a^*}{c}\, K\left(2 {V_2} + 3 i \,  
{V_1}\right)\right.\right. + \nonumber \\
&+&\left. \frac{(b^*)^2}{ac^*}\, \overline{K}\left(2 V_2 +3 i \,  
{V_1}\right)\right] + 
\label{eq:dd} \\
&+&  \left. \frac{\kappa}{4\sqrt{Z}}w\, \left(2 \frac{a}{c^*} g_{03}^*
+ 4 \frac{a^*}{c}g_{12} 
-2i \frac{b}{a^*}\tilde g_{12}^* - 4 \frac{b^2}{ca^*}g_{21}^*\right) 
+ c.c. \right\}\,.\nonumber
\end{eqnarray}
Then, from $d[\chi_+ \eta_-]=0$ we get an extra equation for the other 
function
\begin{eqnarray}
\frac1c  db &-& \frac{b}{c}\, d\log a^*  = i 
\frac{b}{c}Q +
\frac{b}{c}\kappa Z \,\left(1 - i \,J\right) dh 
-\kappa Z \,\overline{K} dh 
+ \nonumber \\
&+& \frac{\kappa}{16 \sqrt{Z}}\left[ 
\frac{a^*}{c}\left(J + 3 i \right)\left(V_1 +
2i\, V_2\right)+\frac{c^*}{a}\left(J + 3 i
\right)\left(V_1 - 2i \,V_2\right)+\right.\nonumber\\
&-& 
\frac{b^2}{a^*c}\left(J + 3 i \right)\left(\overline{V}_1 -
2i\, \overline{V}_2\right) + \nonumber \\
&+&\left.\frac{b^*}{a} \overline{K}\left(2 {V_2} + 3 i \,
{V_1}\right)+\frac{b}{a^*} K\left(2 \overline{V}_{2} + 3 i \,  
\overline{V}_1\right)\right] + 
\label{eq:dc} \\
&-& \frac{\kappa}{4\sqrt{Z}}w \left(4 \frac{b}{a^*} g_{03}^* 
-2i \frac{a^*}{c}\tilde g_{21}
+2i \frac{b^2}{a^*c}\tilde{g}_{12}^*\right)+\nonumber\\
&-&\frac{\kappa}{4\sqrt{Z}}\overline w\left(2 \frac{b}{a^*} g_{12}^*
+ 4\frac{b^*}{a}{g}_{21}- i \left(2\frac{c^*}{a}+ 
\frac{a^*}{c}\right)\tilde{g}_{12}+i \frac{b^2}{a^*c}\tilde g_{21}^*
\right) \,.\nonumber
\end{eqnarray}

Let us then turn to the computation of the $SU(2)$ torsion classes.
Of course, one could express the various components of the 
intrinsic torsion completely in terms of the 3--form flux.
However, we prefer to use where possible also the warp--factor, the
5--form flux and the spinor norms $a$, $b$ and $c$, so as to make the
final expressions a bit more concise.
In some of the following formulae we will also make use of the almost 
complex structure $J = J^1 - \frac{i}{2} w \wedge \overline w$ and of 
the (3,0)--form $\Omega = K \wedge w$, to make more intuitive the 
meaning of some of the various contributions.

The exterior derivative on the globally defined 1--form $w$ is given 
by
\begin{eqnarray}
dw & = & \left[-d\log a^* - d \log c - \frac{1}{4}d \log Z+ 2 i \, \kappa Z 
\, J   dh +2 \kappa\frac{b}{c}Z\, K  dh +
\frac12 \overline \sigma \,\overline w +\right. \nonumber 
\\
&-& \frac{i \kappa}{8\sqrt{Z}}\left(-\frac{c^*}{a}K  
\left(\overline V_1-2i\, \overline V_2\right) 
+ \frac{a^*}{c}K  \left(V_1+ 2i\, V_2\right) + \right. \label{eq:dw} \\
&-&  \left.\left.
2\frac{b}{a^*}\left(V_1-2i\,V_2\right)+ 
\frac{c}{a^*}\overline K \left(V_1+2i\, V_2\right)
+4 i \frac{b}{a^*}(1+i J)  \overline V_2\right)\right] \wedge w\,. \nonumber
\end{eqnarray}
Performing a straightforward calculation, one obtains in 
(\ref{eq:dw}) other terms proportional to the 2--forms 
$J^1$, $K$ and $\overline{K}$.
However, by using (\ref{eq:constdil1}), (\ref{eq:constdil2}) 
and (\ref{eq:gg2}), (\ref{eq:gg3}) one can show that all such terms vanish.
For instance, the $\overline K$ term comes multiplied by $i \, c\,
\tilde g_{21}^* +2 \,b \,g_{30}^*$, which is zero using
(\ref{eq:constdil2}).
The term proportional to the 2--form $K$ comes multiplied by $2 \kappa \frac{b}{c}Z 
\theta^* + \frac{1}{4 \sqrt{Z}}\left(i\frac{b^2}{a^*c} \tilde g_{21}^*- 
i\frac{a^*}{c}\tilde g_{12} + 2 \frac{b}{a^*}g_{12}^*\right)$ and this 
is again vanishing because of (\ref{eq:gg3}), (\ref{eq:gg4}).
There is also a contribution proportional to $J^1$.
This is
$2 Z i\kappa \theta^* + 
i \frac{\kappa}{2\sqrt{Z}}\left(\frac{b^2}{ca^*}g_{30}^*  +  
\frac{a^*}{c} g_{03} - \frac{c}{a^*}g_{12}^*\right)$, which 
vanishes again using the same equations.
Similar simplifications can also be applied to the $w \wedge \overline w$ 
terms to arrive at the final form presented above.

The result in (\ref{eq:dw}) imposes already some strong restrictions on 
the general form of the internal space.
The group structures determine the metric of the internal space.
For an $SU(2)$ structure defined by $J^i$ and $w$
the metric can always be written as
\begin{equation}
ds_6^2 = ds_4^2(y) + w \otimes \overline w\,,
\label{eq:standard}
\end{equation}
where both $ds_4^2$ and $w$ will generically have legs on all possible 
directions of the cotangent space.
However, since from the above result we see that $dw$ is proportional 
to $w$ itself, we can always define the coordinate differentials so 
that  
\begin{equation}
w= e^{A(y)} dy_5 + i \, e^{B(y)} dy_6\,,
\label{eq:wdef}
\end{equation}
for $A$ and $B$ complex functions depending on all the coordinates.
The 4--dimensional part $ds_4^2$ will still generically depend also on
$dy^5$ and $dy^6$.
If, however, the almost product structure $\Pi_a^b$ defined by $w$ and
its dual vector as $\Pi_a^{b} =(w_a \overline{w}^b+ \overline w_a w^b
- \delta_a^b)$ is integrable then the 6--dimensional metric is further 
reduced to a block diagonal form
\begin{equation}
ds_6^2 = \Sigma_{m,n=1}^4 g_{mn}(y) \, dy^m \otimes dy^n + w \otimes
\overline w\,.
\label{eq:product}
\end{equation}
Unfortunately, (\ref{eq:dw}) is not enough to show that $\Pi$ is 
integrable, unless further constraints are imposed.


We can now complete the analysis by computing the intrinsic torsion
contributions coming from the exterior differential on the triplet of
2--forms.
Although we should really compute $dJ^i$, we will present in the 
following $dJ$ and $dK$.
It is then obvious how to extract $dJ^1$.
We just show $dJ$ because it is easier to compute and also  
because it gives direct information on the integrability of the 
associated almost complex structure $J$.
The expression for its exterior differential is
\begin{eqnarray}
dJ & = &   
- J \wedge \left(2 d \log|a| + \frac{1}{2}d \log Z  - 
\kappa Z d h\right) + \nonumber\\
&+& \frac{i}{16\sqrt{Z}}\kappa\left\{\frac{b^*}{a}\left[ 
J \wedge \left(-8 \,\tilde g_{12}\, \overline w - 12 \,\tilde g_{21}\, w- 9 \,V_1 + 5i\,J   
V_1-2i\, V_2+ 6 \,
J   V_2\right)+\phantom{\frac12}\right.\right. \nonumber\\
&-& \left.\phantom{\frac12}4 i\, w \wedge  \bar w \wedge (1-iJ)  \left(V_1 - 2i \,
V_2\right) -
16\, T^1 \wedge {w} - 16 
g_{21}\, K \wedge\overline w\right] + 
\nonumber \\
&+&  i \frac{c^*}{a}\left[-\frac12 w \wedge  \bar w \wedge K   
\left(V_1 -10 i V_2\right) + 8 \,\tilde 
g_{12}  \,K \wedge\overline w + \nonumber \right.\\
&-&\left.\left.\phantom{\frac12} \!\!\!\! J^1 \wedge K \left(2 V_2+5i\, 
V_1\right) - J \wedge \left(16 \,g_{03} \, \overline w + 
24\, g_{12} \, w \right)\right] - c.c.\right\}\,. \label{eq:dJ}
\end{eqnarray}
It is worth noting that although there are in principle non 
vanishing $(3,0)$ or $(0,3)$ components, their coefficient 
vanish identically using (\ref{eq:constdil1}), (\ref{eq:constdil2}).
Finally, we can also compute
\begin{eqnarray}
dK  & = & K \wedge \left[ - d \log (c^*a^*) - i Q - 
\frac{1}{2}d \log Z\right]+ \nonumber\\
&+& \frac{3\kappa}{\sqrt{Z}}\Omega \left[- Z^{3/2}\theta + 
\frac{c}{3a^*}g_{03}^* + \frac{i}{6}\frac{b}{a^*}\tilde g_{12}^* - 
\frac{a}{2c^*}g_{03}^*\right] + \nonumber\\
&+& \frac{3\kappa}{\sqrt{Z}} K \wedge \bar w \left[Z^{3/2}\bar 
\theta + \frac{c}{2a^*}g_{12}^* + \frac{i}{4}\frac{b}{a^*}\tilde 
g_{21}^* - \frac{i}{6}\frac{b^*}{a}\tilde g_{12} - 
\frac{a}{3c^*}g_{12}^*+\frac{(b^*)^2}{3 c^* a}g_{21}\right] + \nonumber\\
&+& \frac{\kappa}{\sqrt{Z}} J \wedge  w \left[\frac{c}{a^*}\tilde 
g_{12}^* - 2 i \frac{b}{a^*}g_{21}^* + i \frac{b^*}{a} g_{12} - 
\frac{a}{2c^*}\tilde g_{12}^* + \frac12 \frac{(b^*)^2}{c^* a} \tilde 
g_{21}\right] + \label{eq:dK}\\
&+& 2i\kappa \frac{b^*}{c^*} Z \, J \wedge (1 - i J)   d h - i \kappa 
\frac{b^*}{c^*} Z \, J \wedge  K   dh + \nonumber \\
&+& \frac{\kappa}{16 \sqrt{Z}} J \wedge \left[-6 \frac{a}{c^*}(1 - i J) 
\overline V_1+6 \frac{(b^*)^2}{ac^*} (1 - i J) V_1 -
4i \frac{a}{c^*}(1 - i J) \overline V_2  + 8 
\frac{c}{a^*}\left(1-iJ\right)\overline V_1 + 
\right.\nonumber  \\
&-& \left. 4i \frac{(b^*)^2}{ac^*} (1 - i J)  V_2 +
8 \frac{b}{a^*}  K  \overline V_1 -2 \frac{b^*}{a} \, K\left(
V_1 - 2 i \, V_2 \right)\right] + \nonumber 
\\
&+& \frac{\kappa}{2\sqrt{Z}} w  \wedge \bar w \wedge  \left[3 Z^{3/2} 
K   dh - \frac{i}{4}\frac{b^*}{a} K   (V_1 +2 i 
V_2) +\frac38 i \frac{b}{a^*} K   (\overline V_1 +2 i 
\overline V_2)+ \right.\nonumber \\
&+&\left.  \left(1-iJ\right) \left(\frac{3}{8}i\,  
\frac{c}{a^*}(\overline V_1 +2 i\, \overline V_2) -\frac{i}{2} 
\frac{a}{c^*}(\overline V_1+2i\,\overline V_2) + \frac{(b^*)^2}{a c^*}\frac{i}{2} 
(V_1 -2 i \,V_2)\right)\right] + \nonumber \\
&-& \frac{\kappa}{\sqrt{Z}} w \wedge \left(\frac{a}{c^*} \overline T^2 - 
\frac{(b^*)^2}{ac^*} T^1\right) + \nonumber \\
&+& \frac{\kappa}{\sqrt{Z}} \frac{c}{a^*}\, \bar w \wedge  \overline 
T^1 + \frac{\kappa}{4 \sqrt{Z}}\frac{c}{a^*} \, J^1 \wedge 
\left(1+iJ\right) \left(\overline V_1+ 2 i\, \overline V_2\right)\,. \nonumber 
\end{eqnarray}
Also this expression has been simplified using the constraints on the 
fluxes coming from the dilatino and external gravitino.
Terms of the form $J \wedge \bar w$, which are present in the straightforward 
calculation, vanish using (\ref{eq:dc}) 
and (\ref{eq:constdil1}), (\ref{eq:constdil2}).

Having the complete form of the intrinsic torsion for the $SU(2)$ 
structure, we can make some comments on the integrability of the complex 
structure $J$.
It is known that given the $SU(3)$ structure defined by $J$ and $\Omega = 
K \wedge w$, the torsion classes which describe the integrability of 
$J$ are ${\cal W}_1$ and ${\cal W}_{2}$, as can be read from
\begin{eqnarray}
d J &=& \frac34 \,i\, \left( {\cal W}_1 \,\overline\Omega -\overline{\cal
W}_1\,\Omega\right)  +{\cal W}_3 +  J \wedge {\cal W}_4 \;, 
\label{dJclass}\\[2mm]
d\Omega &=&  {\cal W}_1  J \wedge  J +  J \wedge  
{\cal W}_2 + \Omega\wedge
{\cal W}_5\,. \label{dpsiclass}
\end{eqnarray}
It is already clear from $dJ$ that ${\cal W}_1 = 0$, 
since there is no $(0,3)$ or $(3,0)$ form in (\ref{eq:dJ}).
Moreover, $dw$ is always proportional to $w$ and therefore
the only obstruction to the integrability of $J$ can only be given by 
the $(1,2)$ contributions coming from $dK$.
By inspection of (\ref{eq:dK}) it is easy to see that only its last line 
contains objects of $(1,2)$ type.
From there we conclude that
\begin{equation}
{\cal W}_2 = \frac{\kappa}{4\sqrt{Z}}\frac{c^*}{a}\left( 8i\, \overline T^1 - 
w \wedge \left(1+iJ\right) \left(\overline V_1+ 2i \overline
V_2\right)\right)\,,
\label{eq:W2}
\end{equation}
and therefore $J$ is an integrable complex structure only if the 
following additional constraints on the 3--form flux are imposed:
\begin{equation}
T^1=0\,, \quad (1-iJ) \left(V_1- 2i\, V_2\right) = 0\,.
\label{eq:complexintegra}
\end{equation}
It should be noted that these constraints are somehow orthogonal to the 
ISD requirement (\ref{eq:ISD}) and imply that if such condition is 
imposed the only primitive flux allowed is $g_{21}$.
Of course, this is just one of the possible almost complex structures 
described by the $SU(2)$ structure.
It is anyway possible to show that for any choice of phase in 
(\ref{eq:Jsu23}) ${\cal W}_1$ remains zero.
For ${\cal W}_2$ we get an involved expression depending on $T_1$, 
$T_2$, $V_1$ and $V_2$, which does not seem to vanish for any choice of 
phase.
As noted also by Frey, in the $c = 0$ limit (\ref{eq:W2}) vanishes, 
i.e. ${\cal W}_1 = {\cal W}_2 = 0$, and therefore the complex structure 
is always integrable.

\section{Solutions}
\label{solutions}

In principle, from the expressions we obtained in the previous 
section for the $SU(2)$ structure, one should be able to provide classes of 
manifolds having the right properties to be supersymmetric 
backgrounds to type IIB supergravity. 
Unfortunately, as it is clear by inspection of
(\ref{eq:dw}), (\ref{eq:dJ}) and (\ref{eq:dK}), although $dw$ and
$dJ^i$ give us some interesting information on the general properties
of the internal manifold, they are not very illuminating on how to
reproduce them in explicit examples.
We leave for the future a more detailed analysis of these conditions 
as well as the presentation of more examples.
Here we will anyway provide some very simple classes of manifolds 
which have the right group structures.
First, we will make contact with the known results obtained for the 
type A (\ref{eq:Aansatz}) and type B Ansatz (\ref{eq:Bansatz}).
Then we will provide a new class of supersymmetric configurations 
which admit also $(0,3)$ flux.

\subsection{Known limits: type B and type A solutions}

In order to make contact with the general solution using the type B 
Ansatz provided in \cite{Grana:2001xn} we should set $b = c = 0$.
Of course, since in this case there is just one spinor on the internal 
manifold, we cannot discuss anymore constraints on the 
$SU(2)$ structure but only on the $SU(3)$ one.
This means that it is not correct to use the form of $P$ and $G$ as in 
(\ref{eq:Pfl}) and (\ref{eq:Gflux}).
We can however assume that form locally, so that we can check our 
previous conditions.
Moreover one should be careful in setting $b=c=0$ in the 
differential conditions derived from the internal gravitino, since 
it would not be consistent, but one should rather 
reconsider (\ref{eq:gravitino}) in the appropriate conventions.

From the dilatino and external gravitino conditions 
one gets that 
\begin{eqnarray}
(1 + iJ) \Pi = p_2 = 0\,, &\quad& g_{03}= g_{30}= \tilde g_{21}= 
\tilde g_{21}= \left(V_1+2i V_2\right)\,,
\label{eq:fluxescond}\\
\sigma = 4 \kappa Z \, \theta\,, &\quad& (1-iJ)\left(\Sigma - 4\kappa 
Z H\right) = 0\,.
\end{eqnarray} 
The first condition means that an holomorphic dilaton is allowed,
whereas its anti-holomorphic part should vanish $P^{(0,1)}=0$.
It is also evident what mentioned in the introduction that such Ansatz 
does not allow for any $(3,0)$ or $(0,3)$ fluxes.
Moreover, the 3--form flux is also constrained so that the
non--primitive $(1,2)_{NP}$ and $(2,1)_{NP}$ parts are vanishing as
well.
Due to reality of both the warp factor $Z$ and the 5--form flux $h$, 
the second equation tells us that 
\begin{equation}
d \log Z = 4 \kappa Z dh\,, \quad \Rightarrow\quad h = -\frac{1}{4\kappa Z}\,,
\label{eq:dlogZdh}
\end{equation}
which is a mentioned feature of all ISD backgrounds.
From the internal gravitino then one gets that
\begin{equation}
2 d \log |a| = -\kappa Z dh\,,\quad  \Rightarrow \quad |a| = Z^{-1/8 }\,.
\label{eq:aZ}
\end{equation}
If we now decompose the function in front of the spinor into norm and 
phase, $a=|a|e^{i\phi_a}$, and use (\ref{eq:aZ}), 
the differential condition on the spinor 
which is left is 
\begin{equation}
\nabla_m \eta_- = \frac{i}{2} \left(Q_m - 2\partial_m\phi_a\right) \eta_-\,,
\label{eq:nablaeta0}
\end{equation}
which is equivalent to (2.19) of \cite{Grana:2001xn}.
This now implies the differential constraints on the $SU(3)$ 
structure, which read
\begin{equation}
dJ = 0 \,, \quad d \Omega = -i\, \Omega \wedge \left(Q-2 \,d\phi_a\right)\,.
\label{eq:almostCy}
\end{equation}
For consistency from (\ref{eq:gravitino}) we also obtain that 
\begin{equation}
G^{(1,2)} = 0\,,
\label{eq:altre}
\end{equation}
which finally leaves only a $(2,1)$ flux and primitive.
Since $Q$ is a $U(1)$ connection, (\ref{eq:almostCy}) implies that the
internal space is a K\"ahler manifold with vanishing first Chern
class, i.e. a Calabi--Yau manifold, but equipped with a metric which
is not the usual Ricci--flat one.
If we want to use the standard Ricci--flat metric, we can  
reabsorb the extra connection in the phase $\phi_a$, by imposing 
\begin{equation}
d\phi_a = -\frac14 \, J   d \log \left(1-|B|^2\right)\,.
\label{eq:dfia}
\end{equation}
This is possible because in (\ref{eq:almostCy}) only the
anti--holomorphic part of $Q$ appears and this latter can be
integrated to $i/2\, \overline \partial \log \left(1-|B|^2\right)$.
Once (\ref{eq:dfia}) is imposed, we obtain the usual Calabi--Yau
conditions, namely having a K\"ahler form and a closed holomorphic
3--form
\begin{equation}
dJ = d\Omega = 0\,.
\label{eq:CY}
\end{equation}
The final result is the known fact that to preserve supersymmetry with 
the B Ansatz (\ref{eq:Bansatz}) one needs 
a $(2,1)_P$ flux and the internal manifold must be conformally Calabi--Yau.

It should be noted that by choosing $c \neq 0$ and $a = 0$ one gets an
anti--holomorphic dilaton, instead of holomorphic, and the flux is
$(1,2)_P$.
Moreover the signs in (\ref{eq:dlogZdh}), (\ref{eq:aZ}) are reversed.

\medskip

For what concerns the type A Ansatz, we will briefly show that we can 
match the conditions coming from the analysis of the common sector of 
all supergravity theories \cite{Cardoso:2002hd,Gauntlett:2003cy}.
For this reason we assume here that the axion is constant, which means 
that $P = \frac12 d\phi$ and $Q = 0$,  that the 5--form flux is 
vanishing, $h = 0$, and that the only non--vanishing 3--form is the NS 
one, which also means that $\overline G = G$.
In the spinor Ansatz we have to impose that $c = 0$ and $b = a^*$, 
which means again that we have to be very careful in taking the limit 
of our analysis, since only an $SU(3)$ structure is allowed.

The dilatino equation kills immediately the $(3,0)$ and $(0,3)$ parts 
of the flux as expected, $g_{30}= g_{03}= 0$.
Then we obtain that the non--primitive $(2,1)+(1,2)$ components of $G$ are 
related to the dilaton as $G^{(2,1)_{NP}+(1,2)_{NP}} = 2 
\frac{\sqrt{Z}}{\kappa}\, J \wedge JP$.
From the external gravitino we obtain the link between warp--factor and 
the same components of the flux, which then translate into a relation 
between dilaton and warp--factor $d\log Z = d \phi$.
Finally, from the internal gravitino we obtain a relation between the 
norm of the spinor and again the $(2,1)_{NP}+(1,2)_{NP}$ parts of the 
flux, $d\log|a| = -\frac18 d\phi$, and the differential conditions on 
the group structures
\begin{eqnarray}
dJ & = & J \wedge d\phi - \kappa * G\,,  \label{eq:dJtypeA}\\
d\Omega & = & -\frac12 \, \Omega \wedge d\phi\,. \label{eq:dOmega}
\end{eqnarray}
In (\ref{eq:dJtypeA}) the dual of the 3--form flux can be reconstructed 
by using the duality properties of the various flux components as 
discussed in the appendix (\ref{eq:dualitypsi})--(\ref{eq:dual21NP}).

\subsection{A simple supersymmetric configuration with $(0,3)$ flux}

We now work out a simple example of a configuration which preserves
${\cal N} =1$ supersymmetry and has a non--vanishing $(0,3)$ component
of the 3--form flux.
To simplify at most the framework in which we work we will assume 
that $b=0$ and that the flux is given only by
\begin{equation}
G = g_{03} \,\overline KÊ\wedge\overline w + g_{12}\, \overline K \wedge  w\,.
\label{eq:Gsimpl}
\end{equation}
Of course, the Hodge decomposition is done with respect to the complex 
structure (\ref{eq:Jdef}).
However, with respect to the general definitions (\ref{eq:su3su2}) one 
sees that the 3--form flux contains a non--vanishing $(0,3)$ flux for 
any phase such that $\cos \phi \neq 0$ and in that case it contains a 
non--vanishing $(3,0)$ part.
We will also assume that the 5--form is vanishing, i.e. $h=0$.
From its Bianchi identity (\ref{eq:dF5}) we get that
\begin{equation}
dF_5 = i\frac{\kappa}{8}\, G \wedge\overline G = 
-i\frac{\kappa}{8}\,\left(|g_{03}|^2 - |g_{12}|^2\right) 
\Omega \wedge \overline \Omega = 0\,,
\label{eq:dF5ex}
\end{equation}
and this implies that 
\begin{equation}
|g_{03}| = |g_{12}|\,.
\label{eq:g03g12}
\end{equation}

The conditions following from the dilatino equation give
\begin{equation}
p_1 = p_2= 0\,, \quad (1+iJ)\Pi = 0  = K   \Pi\,, \quad 
\Leftrightarrow \quad P^{(0,1)}= 0\,, \hbox{ and } \overline w 
\,\lrcorner\; P = 0\,.
\label{eq:dilatex1}
\end{equation}
This means that the dilaton admits a non--trivial $(1,0)$ component on 
the base space, 
and possibly it is holomorphic with respect to the complex coordinates 
defined by the natural complex structure when this latter is integrable.

From the external components of the gravitino, one gets (using the 
fact that $d\log Z$ is real)
\begin{equation}
\sigma = -\frac{2\kappa a}{\sqrt{Z}c^*}\,g_{03}^* = \frac{2\kappa 
c^*}{\sqrt{Z}a}\,g_{12}\,,
\label{eq:sigex}
\end{equation}
and $H=0$ as requires compatibility with our assumptions.
Therefore
\begin{equation}
d\log Z = \frac{2\kappa}{\sqrt{Z}}\left(\frac{c^*}{a}\,g_{12} \, w + 
\frac{c}{a^*} \,g_{12}^* \, \overline w\right)\,.
\label{eq:dlogZex}
\end{equation}

From the internal gravitino we finally get conditions on the 
normalizations of the spinors defining the $SU(2)$ structure:
\begin{eqnarray}
2 d \log|a| & = & \left(\frac14 - \frac12 \left|\frac{c}{a}\right|^2\right)\, 
d\log Z\,, \label{eq:dlogaex}\\
2 d\log |c| & = & \left(\frac14 - \frac12 
\left|\frac{a}{c}\right|^2\right)\, d\log Z\,. \label{eq:dlogdex}
\end{eqnarray}
A possible solution to such conditions is given by choosing $a$ and $c$ 
real and 
\begin{equation}
a = c = Z^{-1/8}\,.
\label{eq:solex1}
\end{equation}
The differential conditions defining the torsion classes are then
\begin{eqnarray}
dw & = & \frac12 \, d \log Z \wedge w\,, \label{eq:dwex}\\[2mm]
dJ^1 & = & 0\,, \label{eq:dJex}\\[2mm]
dK & = & i \, K \wedge Q\,, \label{eq:dKex}
\end{eqnarray}
where we used (\ref{eq:dlogdex}) and (\ref{eq:solex1}), and
the inherited $SU(3)$ structure satisfies $dJ = 0$, $d\Omega = -\frac12 \Omega 
\wedge \left(d\log Z +2i\, Q\right)$.
For such examples the product structure $\Pi_a^{b} =(w_a \overline{w}^b+ \overline w_a w^b
- \delta_a^b)$ is integrable and $Q$ depends only on the coordinates 
of the base.
The total space is therefore a direct product of a Calabi--Yau 2--fold 
and $\mathbb{R}^2$.
Finally, we impose the $G$ Bianchi identity (\ref{eq:dG}).
A solution, which ensures also the equations of motion at least for 
constant dilaton, is given by 
\begin{equation}
 g_{12} = -g_{03}^*= Z^{-1/2}\, f\left(y_5+ i y_6\right)\,,
\label{eq:solbianchi}
\end{equation}
where $f$ is an arbitrary holomorphic function depending only on the 
fiber coordinates.
From (\ref{eq:dJex}) one can see that the almost complex structure is 
actually integrable and that therefore the axion--dilaton $P$ is 
holomorphic in the appropriate complex coordinates defined by $J$.

For the case of a constant dilaton/axion the solution then reads
\begin{eqnarray}
ds_{10}^2 & = & Z^{-1/2}\, \left(dx_0^2-dx_1^2-dx_2^2-dx_3^2\right) - Z^{1/2}
\, ds_{K3}^2(y_{1\ldots 4}) - Z \,\left(dy_5^2 + 
dy_6^2\right)\,,\label{eq:metricex}\\
G & = & Z^{-1/2}\,\left(-f^*(\overline y) \, \overline \Omega + f(y) \, 
K \wedge  w\right)\,, \label{eq:Gex}\\
\epsilon_{10} & = & Z^{-1/8}\, \left(\varepsilon
\otimes \eta_- + \varepsilon^*\otimes \chi_+\right)\,.  
\label{eq:epsex}
\end{eqnarray}
Consistency further imposes that 
\begin{equation}
Z\left(y_5,y_6\right)= \kappa^2\, \left[\int f(y)\, dy + 
\int f^*(\overline y)\, d\overline y\right]^2\,,
\end{equation}
where $y \equiv y_5 + i y_6$.
It is worth noting that the warp factor is a function which cannot be 
consistently defined on a torus and this implies that the form 
presented above is correct only for a non--compact background.
This agrees with the fact that without adding sources one cannot
obtain compact solutions for nontrivial fluxes 
\cite{deWit:1987xg,Maldacena:2000mw}.

\section{Comments}
\label{comments}

The conditions on the fluxes and on the $SU(2)$ structures of the 
internal manifold given here followed simply by the analysis of the 
supersymmetry conditions.
The equations of motion will further restrict the possible 
solutions.
Under quite general assumptions it has been shown in \cite{Giddings:2001yu} 
that for compact backgrounds the allowed 3--form fluxes are of the ISD 
type.
So far only solutions with $(2,1)$ and primitive flux have been given 
when also supersymmetry is required.
From the analysis we have presented, though, it should be possible to 
compactify on 6--manifolds also in the presence of $(0,3)$ and $(1,2)$ 
non--primitive fluxes, still preserving supersymmetry.

When the internal manifold is not compact, no further restriction on 
the Hodge type of the 3--form fluxes follows from the equations 
of motion.
Holographic descriptions of renormalization group flows are in this 
class of solutions and indeed, by the analysis of 
\cite{Grana:2000jj,Aharony:2002hx,Pilch:2004yg}, simple perturbations of the $AdS_5 
\times S^5$ metric by 3--form fluxes generically turn on all possible 
types of the same flux.
The flow presented in \cite{Polchinski:2000uf} is especially interesting. 
There a confining ${\cal N} = 1$ gauge theory is obtained by adding a 
mass perturbation to ${\cal N} = 4$ Yang--Mills theory.
It was argued that such solution could be described on the dual side 
by a type IIB configuration which interpolates between a vacuum with 
only D3 branes and one where the D3 branes are polarized into D5/NS5 branes 
wrapped on a 2--sphere of the internal space.
The supersymmetry conditions for this flow have been studied up to
second order in the perturbation parameter \cite{Grana:2000jj} (see 
also \cite{Grana:2001xn,Aharony:2002hx,Frey:2003sd}.
Again, the standard A, B or C supersymmetry Ans\"atze are not enough to 
describe it and it can be argued that it must then fall into our 
description.
A detailed analysis of the flow is currently under investigation 
\cite{cardoso:xxxx} and it shows that the internal manifold presents an $SU(2)$ 
structure and the solution of the supersymmetry equations involves again 
the D type Ansatz (\ref{eq:Dansatz}).

The possibility of having general types of fluxes compatible with 
supersymmetry is very important in the context of the gauge/gravity 
correspondence.
It has been shown \cite{Gukov:1999ya,Gukov:1999gr,Giddings:2001yu} that the superpotential
of gauge theories dual to IIB strings on Calabi--Yau manifolds in the
presence of 3--form fluxes is given by $W = \int G \wedge \Omega$.
From the field theory point of view then there is no reason for $W$ to 
vanish even for supersymmetric solutions. 
Supersymmetric configurations are only specified by $\partial_I W = 0$, 
where $I$ runs on the moduli of the theory.
This however cannot happen if the standard solutions of type IIB theory 
are used, since the flux contains only $(2,1)$ fluxes and therefore $W = 
0$.
The framework presented here on the other hand allows for  
non--vanishing $(0,3)$ fluxes, so that $G \wedge \Omega \neq 0$ and 
therefore we it should be employed to obtain dual descriptions of 
those gauge theories which allow for non--vanishing $W$.

\smallskip

Recently, in \cite{Pilch:2004yg}, Pilch and Warner have discussed a
class of ``dielectric'' ${\cal N} = 1$ solutions of type IIB
supergravity which include the dual of the Leigh--Strassler flow.
Their approach is similar to the one presented here, though they 
impose from the beginning a definite Ansatz for the metric and 
set to zero the dilaton/axion.
Having a purely holomorphic 2--form potential, it is obvious 
that it may have non--vanishing $(3,0)$ components.
It is therefore interesting to understand how their solution can fit 
into our discussion.
Following the explanations given in the introduction, their spinor
Ansatz must not be of the A, B, or C type and indeed, as they notice,
it does not fit in those schemes.
It can instead be put in the form of the type D Ansatz 
(\ref{eq:Dansatz}).
The projector on the 10--dimensional spinor presented in 
\cite{Pilch:2004yg} is 
\begin{equation}
\frac12 \left(1 + i \Gamma^0 \Gamma^1\Gamma^2\Gamma^3\Gamma^4
(\cos \beta - e^{-i\phi}\sin \beta \, 
\Gamma^7 \Gamma^{10} *)\right) \epsilon = \epsilon\,,
\label{eq:Pilchconst}
\end{equation}
where the star denotes complex conjugation and we used their numbering  
for the $\Gamma$ matrices.
It is a simple exercise to show that (\ref{eq:Dansatz}) satisfies this 
constraint assuming that $a = A(1+\cos \beta)$, $b=0$, $c = 
A e^{-i\phi}\sin \beta$ and $\chi_+ = \gamma^7 \gamma^{10}\eta_+$, for 
a so far undetermined function $A$.
It can also be shown that their choice of complex structure is not the 
standard one built from $\eta$.
We will return elsewhere on the precise embedding of the full solution 
given in \cite{Pilch:2004yg} into the torsion conditions provided in 
this paper as well as on other solutions which fit into our scheme 
\cite{Cascales:2003ew}.


\bigskip \bigskip

\noindent
{\bf Acknowledgments}

\medskip

\noindent
We would like to thank  G.L. Cardoso, G. Curio, D. L\"ust and N. Prezas 
for enlightening discussions.
This work is partially supported by the European Community's Human Potential
Programme under contract HPRN--CT--2000--00131 Quantum Spacetime.


\appendix

\section{Useful formulae}

For the 10--dimensional notations and conventions we follow \cite{Schwarz:1983qr}.
This means that the 10--dimensional $\Gamma$--matrices 
satisfy the Clifford algebra $\{\Gamma^M,\Gamma^N\} = 2 
\, \eta^{MN}$ where $\eta = diag\{+--\ldots-\}$ and the duality 
relation is
\begin{equation}
\Gamma^{A_1\ldots A_p} = - (-1)^{p(p-1)/2} 
\frac{1}{(10-p)!}\epsilon^{A_1\ldots A_{10}} 
\Gamma_{A_{p+1}\ldots A_{10}}\Gamma^{11}\,,
\label{eq:duality10d}
\end{equation}
where we denote by $A_1$,\ldots $A_{10}$ flat 10--dimensional indices.
In a concrete representation one can choose $\Gamma^0$ antisymmetric and imaginary, 
with the others symmetric and imaginary.
$\Gamma^{11}$ is then symmetric and real.
When performing the reduction on the solution, the 10--dimensional $\Gamma$ 
matrices get split as
\begin{equation}
\Gamma^{\alpha} = \gamma^{\alpha} \otimes 1\,, \quad \Gamma^{a} = \gamma^5 
\otimes \gamma^a\,, \quad \Gamma^{11} = \gamma^5 \otimes \gamma^7\,,
\label{eq:10dgamma}
\end{equation}
where we have used the flat indices on Minkowski space--time
$\alpha=0,\ldots,3$ and on the 6--dimensional internal space $ds^2_6$
$a = 1,\ldots,6$.
We are also assuming that $\gamma^5 = i \gamma^{0123}$ and then the 
6--dimensional duality relation is
\begin{equation}
\gamma_{m_1 \ldots m_k}  =  -i \, 
\frac{(-1)^{\frac{k(k-1)}{2}}}{(6-k)!} \, \epsilon_{m_1 \ldots m_6} 
\gamma^{m_{k+1}\ldots m_6} \gamma^7\,.
\label{eq:duality}
\end{equation}
Again, in a concrete representation the 6--dimensional $\gamma$
matrices are antihermitian $(\gamma^{m})^{\dagger} = -\gamma^m$ and purely
real $(\gamma^m)^* = \gamma^m$.

In the text we have assumed that the globally defined spinors defining
the $SU(3)$ and $SU(2)$ structures are chiral and satisfy 
$\eta_\pm^* = \eta_\mp$.
This happens for instance in a concrete representation where
\begin{equation} 
\gamma^7 = \begin{pmatrix}{0 & -i \cr i & 0}\end{pmatrix}\,, \quad 
\eta_+ = \begin{pmatrix}{-i \xi  \cr \xi}\end{pmatrix}
\,, \quad \eta_- = \begin{pmatrix}{i\xi  \cr \xi  }\end{pmatrix}\,,
\end{equation}
for $\xi^* = \xi$ an arbitrary 4--dimensional spinor.
It may be useful to have an explicit representation of the $\gamma$ matrices 
satisfying all the above properties.
One possibility is for instance given by
\begin{eqnarray}
\gamma^1_{18} = \gamma^1_{23} &=&  \gamma^1_{54} = \gamma^1_{76} = 1\,, 
\label{eq:g1}\\
\gamma^2_{41} = \gamma^2_{72} &=&  \gamma^2_{36} = \gamma^2_{58} = 1\,, 
\label{eq:g2}\\
\gamma^3_{17} = \gamma^3_{42} &=&  \gamma^3_{53} = \gamma^3_{68} = 1\,, 
\label{eq:g3}\\
\gamma^4_{31} = \gamma^4_{28} &=&  \gamma^4_{64} = \gamma^4_{57} = 1\,, 
\label{eq:g4}\\
\gamma^5_{25} = \gamma^5_{43} &=& \gamma^5_{61} = \gamma^5_{78} = 1\,, 
\label{eq:g5}\\
\gamma^6_{12} = \gamma^6_{38} &=&  \gamma^6_{65} = \gamma^6_{74} = 1\,.
\label{eq:g6}
\end{eqnarray}
Useful rearrangements when computing the conditions deriving from supersymmetry 
are
\begin{equation}
\eta_{\pm} \eta^{\dagger}_{\pm} = \frac18 \left(1\pm\gamma^7\right) \pm 
\frac{i}{16} J_{mn} \gamma^{mn}\left(1\pm\gamma^7\right)\,,
\label{eq:fierz1}
\end{equation}
and
\begin{equation}
\eta_+ \eta_+^T = \frac{1}{48} {\Omega}_{mnp} \gamma^{mnp}\,, \quad 
\eta_- \eta_-^T = \frac{1}{48} \overline\Omega_{mnp} \gamma^{mnp}\,.
\label{eq:fierz2}
\end{equation}


We list here all the useful relations (transpose is understood 
where needed to build the appropriate tensor combination) for the 
1--forms:
\begin{eqnarray}
\eta_+ \gamma^m \chi_+ = w^m\,, && \chi_+ \gamma^m \eta_+ = -w^m\,,  \label{eq:1f2}\\
\eta_- \gamma^m \chi_- = \overline w^m\,, &  & \chi_- \gamma^m \eta_- = 
-\overline w^m\,, 
\label{eq:1f1}
\end{eqnarray}
2--forms:
\begin{eqnarray}
\eta_-\gamma_{mn}\eta_+ = i J_{mn}\,, &  & \eta_+\gamma_{mn}\eta_- = -i 
J_{mn}\,, \label{eq:Jsp}\\
\eta_+\gamma_{mn}\chi_- = {K}_{mn}\,, &  & 
\eta_-\gamma_{mn}\chi_+ = \overline{K}_{mn}\,,\label{eq:K1}\\
\chi_- \gamma_{mn}\chi_+ = 2 w_{[m}\overline{w}_{n]} - i J_{mn}\,,&  & 
\chi_+ \gamma_{mn}\chi_- = -2 w_{[m}\overline{w}_{n]} + i J_{mn}\,, 
\label{eq:Jww}
\end{eqnarray}
3--forms:
\begin{eqnarray}
\eta_+\gamma_{mnp}\eta_+ = -\Omega_{mnp}\,, &  & \eta_- 
\gamma_{mnp}\eta_- = - \overline{\Omega}_{mnp}\,, \\
\chi_- \gamma_{mnp}\chi_- = -3 {K}_{[mn}\overline w_{p]}\,, &  & \chi_+ 
\gamma_{mnp}\chi_+ = -3 \overline{K}_{[mn}{w}_{p]}\,, \\
\eta_-\gamma_{mnp}\chi_- = 3i J_{[mn}\overline w_{p]}\,, &  & 
\eta_+\gamma_{mnp}\chi_+ = -3i J_{[mn}{w}_{p]}\,, 
\end{eqnarray}
4--forms:
\begin{eqnarray}
\eta_- \gamma_{mnpq}\eta_+ &=&\eta_+ \gamma_{mnpq}\eta_- = -3 J_{[mn}J_{p]q}\,, 
\label{eq:4f}
\end{eqnarray}
It is also interesting to point out the following duality relations 
\begin{eqnarray}
\epsilon_{ijk}{}^{abc} \Omega_{abc} = -6 i \,\Omega_{ijk}\,, &\quad&
\epsilon_{ijk}{}^{abc} \overline{\Omega}_{abc} = 6 i\, \overline{\Omega}_{ijk}\,,
\label{eq:dualitypsi} \\
\epsilon_{abc}{}^{mnp} \overline K_{mn}w_{p} = -6 i \, \overline 
K_{[ab}w_{c]}\,, &\quad& 
\epsilon_{abc}{}^{mnp} K_{mn}\bar w_{p} = 6 i \, K_{[ab}\bar w_{c]}\,, 
\label{eq:dual21}\\
\epsilon_{abc}{}^{mnp} J_{mn}w_{p} =- 6 i \, J_{[ab}w_{c]}\,, &\quad& 
\epsilon_{abc}{}^{mnp} J_{mn}\bar w_{p} = 6 i \, J_{[ab}\bar w_{c]}\,.
\label{eq:dual21NP}
\end{eqnarray}
These equations show that given 
a 3--form, the $(3,0)+(1,2)_P+(2,1)_{NP}$ parts are selfdual whereas 
the $(0,3)+(2,1)_{P}+(1,2)_{NP}$ are anti--selfdual (Here $P$ stands 
for primitive with respect to $J$ and $NP$ for non--primitive).

\subsection{SU(2) structure and Dirac spinors}

A further alternative possibility to define the $SU(2)$ structure is by 
means of a single Dirac spinor
\begin{equation}
\zeta = \frac{1}{\sqrt{2}} \left(\eta_- + \chi_+\right) = 
\frac{1}{\sqrt{2}} \left(1 + \frac12 {w}_m 
\gamma^m\right)\eta_-\,.
\label{eq:zeta}
\end{equation}
This spinor does not have a definite chirality and $\zeta^*$ is 
independent from $\zeta$.
The structure then follows from
\begin{eqnarray}
1 &=& \zeta^\dagger \zeta\,,\label{eq:id}\\
w_m &=& \zeta^\dagger \gamma_m (1 +\gamma^7)\zeta\,, \label{eq:w}\\
J^1_{mn} & = & i \, \zeta^\dagger \gamma_{mn} \zeta\,, \label{eq:J1}\\
w_{[m}\bar w_{n]} & = & \zeta^\dagger \gamma_{mn }\gamma^7 \zeta \,,
\label{eq:wwbar}
\\
K_{mn} & = & -\zeta^\dagger \gamma_{mn}\gamma^7\zeta^*\,. \label{eq:K}
\end{eqnarray}
The other combinations vanish if they cannot be obtained by the above 
ones by complex conjugation.



\providecommand{\href}[2]{#2}

\end{document}